\documentclass[preprint,amsmath,amssymb,aps,showkeys,showpacs]{revtex4}
\usepackage[english]{babel}
\usepackage{graphicx}
\usepackage{graphics}
\usepackage{amsmath}
\usepackage{dcolumn}
\usepackage{amssymb}
\usepackage{bm}

\begin{document}
\title{On an atom with a magnetic quadrupole moment in a rotating frame}
\author{I. C. Fonseca}
\affiliation{Departamento de F\'isica, Universidade Federal da Para\'iba, Caixa Postal 5008, 58051-900, Jo\~ao Pessoa-PB, Brazil.}

\author{K. Bakke}
\email{kbakke@fisica.ufpb.br}
\affiliation{Departamento de F\'isica, Universidade Federal da Para\'iba, Caixa Postal 5008, 58051-900, Jo\~ao Pessoa-PB, Brazil.}

\begin{abstract}

The quantum description of an atom with a magnetic quadrupole moment in the presence of a uniform effective magnetic field is analysed. The atom is also subject to rotation and a scalar potential proportional to the inverse of the radial distance. It is shown that the spectrum of energy is modified, in contrast to the Landau-type levels, and there is a restriction on the possible values of the cyclotron frequency which stems from the influence of the rotation and scalar potential proportional to the inverse of the radial distance. 


\end{abstract}


\maketitle

\section{Introduction}

Rotating effects have been attracted interests in the quantum field theory with studies in Dirac fields \cite{r10}, scalar fields \cite{r8}, the Dirac oscillator \cite{b10} and the Landau quantization \cite{r8a}. In classical physics, Landau and Lifshitz \cite{landau2} showed for a system in a uniformly rotating frame that the line element of the Minkowski spacetime becomes singular at large distances. In nonrelativistic quantum systems, in turn, it has been pointed out in Refs. \cite{r1,r2,r3} that a contribution to the energy levels stems from the coupling between the angular momentum and the angular velocity of the rotating frame. Another quantum effect associated with rotation is the a phase shift that appears in the wave function associated with the coupling between the spin and the the angular velocity of the rotating frame, which is known as the Mashhoon effect \cite{r4}. Other well-known quantum effects are the Sagnac effect \cite{sag,sag1,sag5} and the Aharonov-Carmi geometric phase \cite{ac2} which are related to phase shifts in the wave function of a quantum particle that can be observed in interferometry experiments \cite{r5,r6,r1,r2}. Rotating effects have also been observed in the quantum Hall effect \cite{cond1}, spintronics \cite{spint1,spint2,spint3}, quantum rings \cite{r12,r11,dantas}, Bose-Einstein condensation \cite{cond2} and in the presence of the Kratzer potential \cite{ob}.

Recently, we have studied the effects of rotation on a neutral particles system where a Landau-type quantization \cite{fb5} stems from the interaction of the magnetic quadrupole moment of a neutral particle (atom or molecule) with external fields. Studies of quadrupole moments of atoms and molecules have been made in noncommutative quantum mechanics \cite{nonc}, in single crystals \cite{quad}, nuclear quadrupole interactions \cite{nucquad,nucquad2,quad16}, molecules \cite{quad-1,quad-2,quad3,quad17}, atoms \cite{quad6,prlquad} and superposition of chiral states \cite{quad18}. In this paper, we analyse the effects of rotation and a static scalar potential proportional to the inverse of the radial distance on the Landau-type system associated with a neutral particle with a magnetic quadrupole moment. We show that analytical solutions to the Schr\"odinger equation can be achieved, where the possible values of the cyclotron frequency related to the Landau quantization for a neutral particle with a magnetic quadrupole moment are determined by the angular velocity of the rotating frame, the parameter associated with the static scalar potential proportional to the inverse of the radial distance and the quantum numbers of the system.

The structure of this paper is: in section II, we make a brief review of the Landau-type quantization for an atom with magnetic quadrupole moment, and thus, we analyse the effects of rotation and a scalar potential proportional to the inverse of the radial distance on the system by searching for bound state solutions; in section III, we present our conclusions.

\section{Effects of rotation and a scalar potential}

Let us start by making a brief review of the quantum description of a neutral particle (molecule or atom) with a magnetic quadrupole moment in a rotating frame. Let us assume that the system rotates with a constant angular velocity $\vec{\Omega}=\Omega\,\hat{z}$ (where $\hat{z}$ is a unit vector in the $z$-direction), then, as pointed out in Refs. \cite{landau3,landau4,dantas,anan,r13}, the general form of the Schr\"odinger equation is given by (with the units $\hbar=c=1$)
\begin{eqnarray}
i\,\frac{\partial\psi}{\partial t}= \hat{H}_{0}\,\psi-\vec{\Omega}\cdot\hat{L}\,\psi,
\label{1}
\end{eqnarray} 
where $\hat{H}_{0}$ is Hamiltonian operator in the absence of rotation, $\hat{L}$ is the angular momentum operator given by $\hat{L}=\vec{r}\times\hat{\pi}$. For a nonrelativistic neutral particle (atom or molecule) with a magnetic quadrupole moment, the Hamiltonian operator is given by $\hat{H}_{0}=\frac{\hat{\pi}^{2}}{2m}-\vec{M}\cdot\vec{B}+V$ \cite{fb4,fb5}, where $m$ is the mass of the particle, $V$ is a scalar potential and $\hat{\pi}=\hat{p}-\,\vec{M}\times\vec{E}$. Note that the fields $\vec{E}$ and $\vec{B}$ are the electric and magnetic fields in the laboratory frame and the components of the vector $\vec{M}$ are defined by $M_{i}=\sum_{j}M_{ij}\,\partial_{j}$, where $M_{ij}$ is a symmetric and a traceless tensor called as the magnetic quadrupole moment tensor \cite{pra,prc}.

Let us now consider a particular case where the magnetic quadrupole moment has only two non-null components given by $M_{\rho z}=M_{z\rho}=M$, where $M$ is a constant $\left(M>0\right)$ and interacts with the electric field $\vec{E}=\frac{\lambda\,\rho^{2}}{2}\,\hat{\rho}$, where $\lambda$ is a constant associated with a nonuniform distribution of electric charges inside a non-conductor cylinder, $\rho=\sqrt{x^{2}+y^{2}}$ is the radial coordinate and $\hat{\rho}$ is a unit vector in the radial direction. As shown in Ref. \cite{fb2}, this interaction between the magnetic quadrupole moment and the electric field gives rise to an analogue of the Landau quantization in the sense that the neutral particle is placed in a region with a uniform effective magnetic field defined as $\vec{B}_{\mathrm{eff}}=\vec{\nabla}\times\left[\vec{M}\times\vec{E}\right]=\lambda\,M\,\hat{z}$. Recently, we have analysed the effects of rotation of this Landau-type system \cite{fb5}. Our focus in this work is on the effects of a static scalar potential proportional to the inverse of the radial distance and rotation on an atom with a magnetic quadrupole moment in the presence of the uniform effective magnetic field $\vec{B}_{\mathrm{eff}}$. It is worth emphasizing that scalar potentials proportional to the inverse of the radial distance have interests in studies of position-dependent mass systems \cite{pdm2,pdm3,pdm5}, molecules \cite{molecule,ct5,ct6}, the Kratzer potential \cite{kratzer,kratzer2,kratzer3}, propagation of gravitational waves \cite{ct14} and quark models \cite{quark}. Other interesting studies have been made in Refs. \cite{extra,extra4,extra2,extra3}. Thereby, let us consider a scalar potential proportional to the inverse of the radial distance $V\left(\rho\right)=\frac{\vartheta}{\rho}$, where $\vartheta$ is a constant parameter that characterizes the scalar potential \cite{quark}. The Schr\"odinger equation (\ref{1}) becomes:
\begin{eqnarray}
i\frac{\partial\psi}{\partial t}&=&-\frac{1}{2m}\left[\frac{\partial^{2}}{\partial\rho^{2}}+\frac{1}{\rho}\,\frac{\partial}{\partial\rho}+\frac{1}{\rho^{2}}\,\frac{\partial^{2}}{\partial\varphi^{2}}+\frac{\partial^{2}}{\partial z^{2}}\right]\psi+i\frac{M\,\lambda}{m}\,\frac{\partial\psi}{\partial\varphi}+\frac{M^{2}\,\lambda^{2}}{2m}\,\rho^{2}\,\psi\nonumber\\
[-2mm]\label{1.6}\\[-2mm]
&+&i\Omega\,\frac{\partial\psi}{\partial\varphi}+M\,\lambda\,\Omega\,\rho^{2}\psi+\frac{\vartheta}{\rho}\,\psi.\nonumber
\end{eqnarray}

By writing the wave function in the form $\psi\left(t,\,\rho,\,\varphi,\,z\right)=\Psi\left(t\right)\,\Phi\left(\varphi\right)\,Z\left(z\right)\,F\left(\rho\right)$, then, we obtain $\Psi\left(t\right)=e^{-i\mathcal{E}t}$, $\Phi\left(\varphi\right)=e^{il\varphi}$, $Z\left(z\right)=e^{ip_{z}\,z}$, where $l=0,\pm1,\pm2,\ldots$ and $p_{z}$ is a constant. Henceforth, we assume that $p_{z}=0$ in order to reduce the system to a planar system. By performing a change of variables given by $r=\sqrt{\delta}\,\rho$, hence, the Schr\"odinger equation (\ref{1.6}) becomes
\begin{eqnarray}
\left[\frac{d^{2}}{dr^{2}}+\frac{1}{r}\frac{d}{dr}-\frac{l^{2}}{r^{2}}\,-r^{2}\,-\frac{2m\vartheta}{\sqrt{\delta}\,r}\,+\frac{\Theta}{\delta}\right]F=0,
\label{1.8}
\end{eqnarray}
where we have defined the parameters in Eq. (\ref{1.8}):
\begin{eqnarray}
\delta^{2}&=&\frac{m^{2}\omega^{2}}{4}+m^{2}\Omega\,\omega;\nonumber\\
\Theta&=&2m\mathcal{E}+2m\,\Omega\,l+m\omega\,l;\label{1.9}\\
\omega&=&\frac{2\,M\,\lambda}{m},\nonumber
\end{eqnarray}
where the parameter $\omega$ is the cyclotron frequency of the Landau quantization associated with an atom with a magnetic quadrupole moment obtained in Ref. \cite{fb2}.

A possible way of writing the solution to Eq. (\ref{1.8}) is to consider the radial wave function to be well-behaved at the origin and vanishes at $r\rightarrow\infty$. This permit us to write the function $F\left(r\right)$ in terms of an unknown function $h\left(r\right)$ as follows \cite{mhv,vercin,heun,fb7}:
\begin{eqnarray}
F\left(r\right)=e^{-\frac{r^{2}}{2}}\,r^{\left|l\right|}\,h\left(r\right).
\label{1.10}
\end{eqnarray}
By substituting Eq. (\ref{1.10}) into Eq. (\ref{1.8}), we obtain the following second order differential equation:
\begin{eqnarray}
\frac{d^{2}h}{dr^{2}}+\left[\frac{2\left|l\right|+1}{r}-2r\right]\frac{dh}{dr}+\left[\frac{\Theta}{\delta}-2-2\left|l\right|-\frac{2m\vartheta}{r\,\sqrt{\delta}}\right]h=0;
 \label{1.11}
\end{eqnarray}
thus, the function $h\left(r\right)$, which is the solution to Eq. (\ref{1.11}), is called in the literature as the biconfluent Heun function \cite{heun}: $h\left(r\right)=H_{B}\left(2\left|l\right|,\,0,\,\frac{\Theta}{\delta},\,\frac{4m\vartheta}{\sqrt{\delta}},\,r\right)$. Let us write the solution to Eq. (\ref{1.11}) as a power series expansion around the origin: $h\left(r\right)=\sum_{k=0}^{\infty}c_{k}\,r^{k}$ \cite{arf}; thus, after some calculations, we obtain the relation
\begin{eqnarray}
c_{1}=\frac{2m\vartheta}{\sqrt{\delta}\left(1+2\left|l\right|\right)}\,c_{0},
\label{1.13a}
\end{eqnarray}
and the recurrence relation: 
\begin{eqnarray}
c_{k+2}=\frac{2m\vartheta}{\sqrt{\delta}\left(k+2\right)\left(k+2+2\left|l\right|\right)}\,c_{k+1}-\frac{\frac{\Theta}{\delta}-2-2\left|l\right|-2k}{\left(k+2\right)\left(k+2+2\left|l\right|\right)}\,c_{k}.
\label{1.13}
\end{eqnarray}

By starting with $c_{0}=1$, then, Eqs. (\ref{1.13a}) and (\ref{1.13}) allow us to obtain other coefficients of the power series expansion, for instance, 
\begin{eqnarray}
c_{1}=\frac{2m\vartheta}{\sqrt{\delta}\left(1+2\left|l\right|\right)};\,\,\,\,\,\,\,
c_{2}=\frac{4m^{2}\vartheta^{2}}{2\,\delta\left(2+2\left|l\right|\right)\left(1+2\left|l\right|\right)}-\frac{\left(\frac{\Theta}{\delta}-2-2\left|l\right|\right)}{2\left(2+2\left|l\right|\right)}.
\label{1.14}
\end{eqnarray}

In search of bound state solutions, we wish to obtain polynomial solutions to Eq. (\ref{1.11}), therefore, we must impose that the biconfluent Heun series becomes a polynomial of degree $n$. This happens when
\begin{eqnarray}
\frac{\Theta}{\delta}-2-2\left|l\right|=2n;\,\,\,\,\,c_{n+1}=0,
\label{1.15}
\end{eqnarray}
where $n=1,2,3,\ldots$. The first condition given in Eq. (\ref{1.15}) yields
\begin{eqnarray}
\mathcal{E}_{n,\,l}=\sqrt{\frac{\omega^{2}}{4}+\Omega\,\omega}\,\,\left[n+\left|l\right|+1\right]-\frac{1}{2}\,\omega\,l-\Omega\,l,
\label{1.16}
\end{eqnarray}
where $n$ is the quantum number related to radial modes, $l$ is the angular momentum quantum number and the last term of the right-hand side of Eq. (\ref{1.16}) is the Page-Werner {\it et al} term \cite{r1,r2,r3}, i.e., the coupling between the angular momentum and the angular velocity of the rotating frame. 

On the other hand, the second condition established in Eq. (\ref{1.15}) requires the choice of a parameter of the system in which some appropriate values of it can be chosen in the laboratory. For instance, we can select the cyclotron frequency $\omega$ defined in Eq. (\ref{1.9}), since the intensity of the electric field can be chosen in the laboratory through the parameter $\lambda$ associated with the volume charge density. Therefore, this choice of the cyclotron frequency shows us that a polynomial solution to the function $h\left(r\right)$ is obtained because both conditions of Eq. (\ref{1.15}) are satisfied. This can be clarified by considering the lowest energy state of the system $n=1$. In the lowest energy state, the condition $c_{n+1}=0$ yields $c_{2}=0$, then, by using Eq. (\ref{1.14}), we obtain that the possible values of the cyclotron frequency associated with the lowest energy state of the system are 
\begin{eqnarray}
\omega_{1,\,l}=-2\Omega\pm2\Omega\sqrt{1+\frac{4m^{2}\vartheta^{4}}{\Omega^{2}\left(1+2\left|l\right|\right)^{2}}}.
\label{1.19}
\end{eqnarray}

Note that the possible values of $\omega_{1,\,l}$ given in Eq. (\ref{1.19}) yields $\delta>0$, and thus the asymptotic behaviour of the radial wave function when $r\rightarrow\infty$ is satisfied. Hence, we have in Eq. (\ref{1.19}) that the possible values of the cyclotron frequency associated with the lowest energy state of the system that satisfy the condition $c_{n+1}=0$ are determined by the parameter that characterizes the scalar potential proportional to the inverse of the radial distance, the angular velocity of the rotating frame and the quantum numbers $\left\{n,\,l\right\}$ of the system. Due to this dependence of the quantum numbers of the system, we have labelled $\omega=\omega_{n,\,l}$ in Eq. (\ref{1.19}). Further, by substituting (\ref{1.19}) into Eq. (\ref{1.16}), we have 
\begin{eqnarray}
\mathcal{E}_{1,\,l}=\frac{2m\,\vartheta^{2}\left(\left|l\right|+2\right)}{\left(1+2\left|l\right|\right)}\mp\Omega\,l\,\sqrt{1+\frac{4m^{2}\vartheta^{4}}{\Omega^{2}\left(1+2\left|l\right|\right)^{2}}},
\label{1.20}
\end{eqnarray}
which corresponds to the allowed energies for the lowest energy state of the system. Moreover, the radial wave function (\ref{1.10}) associated with the lowest energy state is given by
\begin{eqnarray}
F_{1,\,l}\left(r\right)=e^{-\frac{r^{2}}{2}}\,r^{\left|l\right|}\left[1+\frac{2m\vartheta}{\sqrt{\delta}\left(1+2\left|l\right|\right)}\,r\right].
\label{1.21}
\end{eqnarray}

For the first excited state $\left(n=2\right)$, the possible values of the angular frequency are:
\begin{eqnarray}
\omega_{2,\,l}=-2\Omega\pm2\Omega\sqrt{1+\frac{m^{2}\,\vartheta^{4}}{\Omega^{2}\left(3+4\left|l\right|\right)^{2}}},
\label{eq:}
\end{eqnarray}
and the allowed energies are given by
\begin{eqnarray}
\mathcal{E}_{2,\,l}=\frac{m\,\vartheta^{2}\left(3+\left|l\right|\right)}{\left(3+4\left|l\right|\right)}\mp\Omega\,l\sqrt{1+\frac{m^{2}\vartheta^{4}}{\Omega^{2}\left(3+4\left|l\right|\right)^{2}}}.
\label{eq:}
\end{eqnarray}

Hence, in order to write the energy levels in a general form, let us use $\omega=\omega_{n,\,l}$, and then
\begin{eqnarray}
\mathcal{E}_{n,\,l}=\sqrt{\frac{\omega^{2}_{n\,l}}{4}+\Omega\,\omega_{n,\,l}}\,\,\left[n+\left|l\right|+1\right]-\frac{1}{2}\,\omega_{n,\,l}\,l-\Omega\,l.
\label{1.22}
\end{eqnarray}

Hence, Eqs. (\ref{1.19}), (\ref{1.20}) and (\ref{1.22}) show us that the spectrum of energy is modified by comparing with the Landau-type levels \cite{fb2} due to the influence of rotation and the scalar potential proportional to the inverse of the radial distance. In contrast to the results of Ref. \cite{fb2}, we have a new expression for the energy levels given in Eq. (\ref{1.22}), where the lowest energy state of the system is determined by the quantum number $n=1$. The cyclotron frequency $\omega=\frac{2M\lambda}{m}$ is replaced with an an angular frequency given by $\varpi=\sqrt{\frac{\omega^{2}}{4}+\Omega\,\omega}$ which arises from the effects of the rotation \cite{fb5}. Besides, the Page-Werner {\it et al} term \cite{r1,r2,r3} is present in the general expression of the energy levels of the system, which corresponds to the coupling between the angular momentum quantum number $l$ and the angular velocity $\Omega$. Observe that the possible values of the cyclotron frequency and the energy associated with the lowest energy state differ from that obtained in Refs. \cite{fb2,fb5,fb7}. By taking $\Omega\rightarrow0$ in Eq. (\ref{1.16}), we recover the behaviour of the Landau-type system under the influence of a Coulomb-type interaction discussed in Ref. \cite{fb7}. By taking $\vartheta=0$ in Eq. (\ref{1.8}), then, we shall recover the analogue of the Landau levels in the rotating frame \cite{fb5}. With $\vartheta=0$ and $\Omega\rightarrow0$, thus, from in Eq. (\ref{1.8}), we recover the analogue of the Landau levels for a neutral particle with a magnetic quadrupole moment \cite{fb2}.

Finally, in our search for polynomial solutions to the function $h\left(r\right)$, we have analysed the lowest energy state of the system and seen that only some specific values of the cyclotron frequency $\omega$ are allowed, where the possible values are determined by the parameter that characterizes the scalar potential, the angular velocity of the rotating frame and the quantum numbers $\left\{n,\,l\right\}$ of the system. The same analysis can be made for other energy levels, and thus new expressions for the cyclotron frequency can be obtained.

\section{conclusions}

We have discussed the behaviour of a neutral particle with a magnetic quadrupole moment in a region with a uniform effective magnetic field under the effects of rotation and a scalar potential proportional to the inverse of the radial distance. We have seen that rotating effects changes the angular frequency by comparing with the cyclotron frequency associated with the analogue of the Landau levels by yielding a new contribution to it. Moreover, rotating effects have gave rise to the Page-Werner {\it et al} term \cite{r1,r2,r3} in the spectrum of energy of the system. In the search for polynomial solutions to the radial wave function, we have analysed the lowest energy state of the system and observed that there exists a restriction on the possible values of the angular frequency of the system, where these values are determined by the angular velocity of the rotating frame, the parameter associated with the scalar potential proportional to the inverse of the radial distance and the quantum numbers of the system.

An interesting discussion about a a neutral particle with a magnetic quadrupole moment in a region with a uniform effective magnetic field under the effects of rotation and a scalar potential proportional to the inverse of the radial distance is in the presence of a disclination and a screw dislocation. It is well-known that disclinations and dislocations are topological defects \cite{kleinert,kat,put} and their influence on quantum systems have been reported in the literature in studies of the Kaluza-Klein theory \cite{fur4}, the Landau quantization \cite{fur,fil}, the quantum Hall effect \cite{hall}, quantum scattering \cite{valdir2,fur3,fur2}, spin currents \cite{spin}, cylindrical shells \cite{shell} and geometric quantum phases \cite{fur6,fur7}.

\acknowledgments

The authors would like to thank the Brazilian agencies CNPq and CAPES for financial support.

\end{document}